\documentclass[10pt,letterpaper,twocolumn]{article}
\usepackage{ol}
\usepackage[latin1]{inputenc}
\usepackage{amsmath}
\usepackage{amssymb}
\usepackage{latexsym}
\usepackage{graphicx}
\usepackage{textcomp}

\newcommand{\laplace}{\mathrm \Delta}

\begin{document}
\twocolumn[%

\title{Counterpropagating dipole-mode vector soliton}

\author{Jochen Schr{\"o}der and Philip Jander and Cornelia Denz}
\affiliation{Institute of Applied Physics, Westf\"{a}lische
Wilhelms-Universit\"{a}t M\"{u}nster, \\ D-48149 M\"{u}nster,
Germany}
\author{Tobias Richter and Kristian Motzek and Friedemann Kaiser}
\affiliation{Institute of Applied Physics, Darmstadt University of
Technology,\\
D-64289 Darmstadt, Germany}

\begin{abstract}
  We experimentally observe a counterpropagating dipole-mode
  vector soliton in a photorefractive SBN:60Ce crystal. We investigate the
  transient formation dynamics and show that the formation process differs
  significantly from the copropagating geometry. The experimental results are
  compared with fully anisotropic numerical simulations which show good
  qualitative agreement.
\end{abstract}
]

Spatial optical solitons in a counterpropagating (CP) geometry
have been proposed some time ago \cite{haelterman_opt_comm}. It
has not been until recently however, that they have drawn wider
attention
\cite{cohen_cp,cohen_cp_exp,cp_interaction_segev,belic_jander_cp}
and were investigated theoretically in saturable Kerr and local
photorefractive media in one transverse dimension (1D). In
addition, Cohen et al. experimentally demonstrated the existence
of stable CP solitons with narrow stripe beams
\cite{cohen_cp_exp}.

In a previous publication, we numerically predicted the existence
of stable two-dimensional higher-order counterpropagating vector
solitons in saturable Kerr media \cite{dynamic_cp_pre_motzek},
commonly used as an approximation for the photorefractive
nonlinearity. In the present paper we will experimentally and
numerically demonstrate the existence of a counterpropagating
dipole-mode vector soliton in a photorefractive SBN:60Ce crystal
and investigate the transient formation dynamics.

A dipole-mode vector soliton consists of two mutually incoherent
beams: an optical dipole and a fundamental-mode(FM) beam. The
individually propagating dipole does not form a spatial soliton
due to repulsion of the dipole components
\cite{dipole_observation,carmon1}. However, if an FM beam which is
incoherent to the dipole is launched in between the dipole
components, they will be ``trapped'' via incoherent attraction
\cite{kivshar_dipole,dipole_observation}.

The counterpropagating geometry poses some additional challenges
to the experimental setup. A soliton propagating inside a
photorefractive crystal will be displaced in the positive
direction of the crystal $c$-axis. This effect is well-known as
the so-called ``beam-bending''-effect caused by charge-carrier
diffusion. This effect has been extensively investigated in the
case of single soliton formation. For the case of interaction
copropating solitons or copropagating vector solitons, the effect
does not change the fundamental interaction behaviour as the beams
are displaced by the same amount \cite{transient_dynamics_denz}.
Two counterpropagating solitons however are bend in the same
direction. Thus when launched in a head-on configuration they will
 cross inside the crystal, resulting in a change of the
soliton interaction length. Therefore, beam-bending has a
significant effect on the transient formation dynamics as will
become apparent later. Although at a first, rough glance the
counterpropagating geometry might be very similar to
configurations for e.g. double phaseconjugate (DPC) mirrors, the
actual experiments are quite distinct. Both interacting beams are
mutually incoherent, thereby not interfering in the material. The
lack of an interference grating is obvious due to the
insensitivity of the beams to slight angular changes once the
solitons are formed. This effect can directly be seen when
observing the dynamics of the soliton formation. In contrast, the
existence of a grating would result in a highly selective
Bragg-condition. Such a behaviour is typical for double phase
conjugation configurations, but does not appear in our case.
Moreover, the scales are much smaller than for DPC experiments
(input beams are in the range of 20 micrometers, interaction
angles are smaller than $1^{\circ}$), and effects that are
typically associated with four-wave mixing and phase-conjugation
do not appear on that scale. Nevertheless, in all our experiments
we checked thoroughly that the effects we are observing are indeed
caused by counterpropagating solitons.
\begin{figure}
  \begin{center}
      \includegraphics{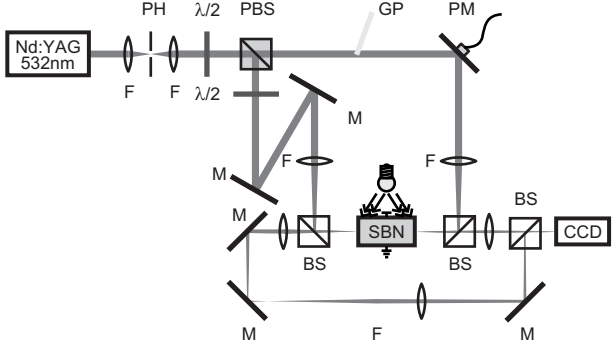}
  \end{center}
  \caption{Experimental setup for the realization of a counterpropagating
  dipole-mode vector soliton. (PH:
  pinhole, F: lens, $\lambda/2$: waveplate, (P)BS: (polarizing) beamsplitter,
  GP: glass-plate, M: mirror, PM: piezo-driven mirror)}
  \label{fig:setup}
\end{figure}

The experimental setup is depicted in Fig.~\ref{fig:setup}. A
frequency-doubled Nd:YAG-laser beam is divided into two arms. The
beams are rendered mutually incoherent by a means of a
piezo-driven mirror (PM) oscillating much faster than the response
time of the crystal and are then focused onto the two faces of the
photorefractive SBN-crystal via two beamsplitters. Due to the
mutual incoherence of the beams, they do not interfere in the
material nor produce an interference grating, thereby only
interacting via their common phase modulation and induced
refractive index channel. The crystal is biased with an external
DC field along the $x$-axis parallel to the crystallographic
$c$-axis. To take advantage of the large $r_{33}$ electro-optic
coefficient of SBN, the beams are polarized along the same
direction. Background illumination is realized with a white light
source, which allows adjustment of the dark-intensity. To be able
to observe the two crystal faces simultaneously, both are imaged
onto the same CCD-camera. The optical dipole is generated by
entering a tilted glass-plate into one of the arms, a well-known
technique for the creation of optical dipoles, and is oriented
perpendicular to the c-axis to minimize effects of the
photorefractive anisotropy\cite{incoherent_interaction}. In order
to compensate for the above-mentioned beam-bending, the beams are
launched into the crystal in such a way that the input face of one
beam coincides with the exit face of the counterpropagating beam
propagating as an individual soliton. This results in a small
angle ($<1^{\circ}$) between the two incident beams. The power of
the FM beam is $P_{FM-beam}\approx 1.00~\mu W$ and the power of
the dipole $P_{dipole}\approx 1.35~\mu W$. The approximate
diameter of the beams at the input face is $d\approx 20~\mu m$
FWHM. The geometry of the SBN crystal is $23(a)\times 5(b)\times
5(c)~mm$ and propagation is along the $a$-axis of the crystal for
maximum propagation length, corresponding to approximately 4
diffraction lengths. The external field was set to $E_{ext}\approx
1.2~kV/cm$ and the background-illumination was adjusted in order
to allow stable self-focusing and soliton formation of the
individually propagating FM beam.

\begin{figure}
  \begin{center}
      \includegraphics{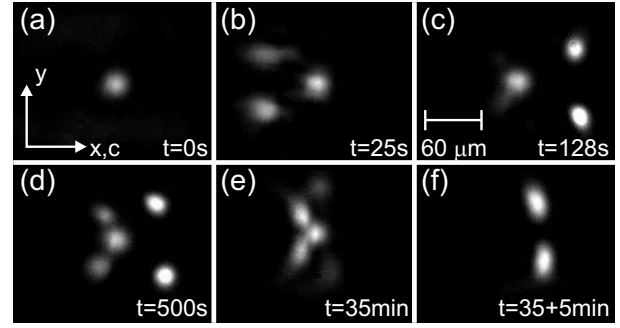}
  \end{center}
  \caption{Images of the dipole exit face of the crystal(beam
  from FM input face to camera is blocked). The spot in the
  middle is the reflection of the FM input serving as a reference.
  (a) linear propagation. Soliton formation at times $t=25~s$ (b), $t=128~s$ (c) and
  $t=500~s$(d) after the nonlinearity has been turned on.
  (e) stable state after $t\approx 35~min$. (f) individually propagating
  dipole
  approximately $t=5~min$ after the FM beam was blocked.}
  \label{fig:dipole-entstehung}
\end{figure}
The formation of the dipole-mode vector soliton is shown in
Fig.~\ref{fig:dipole-entstehung}. The images are taken at the exit
face of the dipole beam, the less intense spot in the middle of
the images is the reflection of the FM input at the crystal face.
It serves as a reference and does not influence the soliton
formation process. When the nonlinearity is turned on, the dipole
initially self-focuses to the left of the FM beam (Image (b)).
This is due to the initial angle between the beams. The dipole is
then strongly deflected to the right, as beam-bending and
attraction from the FM soliton act in the same direction. The
dipole splits up and two spots of low intensity with a much
smaller separation appear on the left side of the FM beam (c). As
time elapses, this trapped part of the dipole draws intensity from
the non-trapped part and the non-trapped part is attracted
horizontally toward the FM input (d). After approximately 35
minutes, the system reaches equilibrium, with almost all power of
the dipole trapped (e). When the FM beam is blocked, the dipole
components quickly repel and move to the right due to
beam-bending.  Image (f) shows the dipole after propagating
individually for approximately 5 minutes. The separation of the
dipole components of the trapped and non-trapped part in image (e)
was $d_{trapped}\approx 38~\mu m$ and $d_{non-trapped}\approx
87~\mu m$ respectively, compared to $d_{individual}\approx 57~\mu
m$ for the individually propagting dipole.

The transient formation dynamics differ significantly from the
copropagating case, which makes them particularly interesting. The
most striking fact is the splitting of the dipole into two parts,
which is not observed in the copropagating geometry. Although at a
quick glance such an effect might be attributed to phenomena as
conical scattering, a thourough analysis shows immediately that
those effects can be excluded: the two interacting solitons are
mutually incoherent, and thereby not able to interfere with each
other. Moreover, the angles at which these new beams appear are in
the range of one degree, much smaller than the scattering cone
which should be well above $10^{\circ}$. Instead, the splitting
can be explained by the angle between the beams which is
introduced to compensate effects of beam-bending. Soliton
formation in photorefractive crystals is a two-step process, in
which the beam self-focuses initially and is then displaced by
beam-bending \cite{transient_dynamics_denz}. After self-focusing
of the counterpropagating beams self-focus but before beam-bending
takes place, the beams only cross at one point inside the crystal
due to the angular adjustment. At this point the colliding beams
will interact strongly, which causes the observed splitting. The
non-trapped part of the dipole is then deflected by the combined
forces of beam-bending and incoherent attraction of the dipole,
causing the dipole to ``overshoot''. This explains why the
non-trapped dipole part is initially displaced further then the
individually propagating dipole, which can be seen by comparing
images (c) and (d) with image (f) in
Fig.~\ref{fig:dipole-entstehung}. The second remarkable feature of
the counterpropagating dipole is the long timescale of the
formation process. It differs significantly from copropagating
solitons, with formation times in the range of $20-50$ seconds or
even shorter for higher intensities\cite{transient_dynamics_denz},
compared to around 30 minutes for the counterpropagating
dipole-mode vector soliton to reach a stable state for similar
parameters.
\begin{figure}
  \begin{center}
      \includegraphics{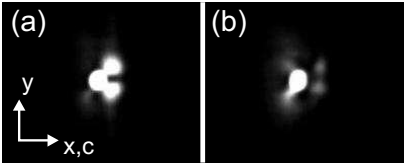}
  \end{center}
  \caption{Images of the dipole input face at steady state $\approx35~min$
  after the nonlinearity has been turned on. (a) The output of the FM beam is
  displaced to the left of the dipole input. (b) Output of the FM when
  the dipole is shortly blocked, a small fraction of the FM beam has coupled
  into the dipole waveguide.}
  \label{fig:gauss}
\end{figure}
As can be seen from Fig.~\ref{fig:dipole-entstehung}(c)-(e), the trapped dipole
is not aligned with the input of the counterpropagating beam, but is displaced slightly to
the left. The output of the FM beam, which can be seen in
Fig.~\ref{fig:gauss} shows a similar picture. The FM
soliton is displaced to the
left of the dipole input. This is in agreement with earlier simulations
\cite{dynamic_cp_pre_motzek} which predict a threshold propagation length where
the symmetry along the dipole axis is broken.
Thus our experimental results are above
this threshold length. Additionally we observe that the the FM beam
splits as well. Fig.~\ref{fig:gauss}(b) shows the exit face of the FM beam
when the dipole is shortly blocked. A fraction of the beam has split from
the main part and has coupled into the dipole waveguide.

\begin{figure}
  \begin{center}
      \includegraphics{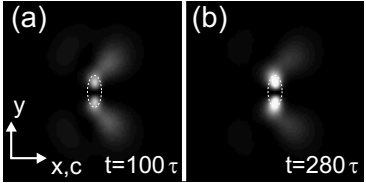}
  \end{center}
  \caption{Numerical simulations of the dipole-mode vector soliton. The images
  show the exit face of the dipole at (a) $t=100~\tau$ and (b)$t=280~\tau$
  after the start of the simulation.
  The location of the input beams is indicated by the dashed
  lines.}
  \label{fig:dipole-numerics}
\end{figure}
To verify our experimental results, we carried out simulations in the
anisotropic, nonlocal model \cite{zozulya_anderson_aniso}.
The two counterpropagating beams are denoted by their slowly varying envelopes $F$ and $B$.
The total optical field can then be written as $F\exp(ikz) + B\exp(-ikz) $, with $k$ being the wavevector
in the undisturbed crystal. Since we are considering $F$ and $B$ to be mutually incoherent, the total intensity
is just given by $I = |F|^2 + |B|^2$. The Kukhtarev model
of the photorefractive crystal leads under simplifying but well justified assumptions to the following equation
\begin{multline}
\frac{\tau}{1+I}\partial_t\laplace\phi +
\laplace\phi + \nabla\ln(1+I)\nabla\phi =  \\
E_{ext} \partial_x\ln(1+I) \\
+ \kappa\left[\laplace\ln(1+I) + (\nabla\ln(1+I))^2 \right]   \label{eqn:pot_eqn}
\end{multline}
for the potential $\phi$ of the electrical screening field, which causes the nonlinear change of the refractive index
via the Pockels effect. $E_{ext}$ denotes the externally applied electrical field, $\kappa = -k_B T/e$ is the diffusive
coupling strength, $\tau$ is the effective time constant of the crystal, and $I$ is measured in units of the background
intensity.

The propagation of the beams is in paraxial approximation described by the set of equations
\begin{subequations}
\label{eqns:cpb_prop_eqns}
\begin{align}
    \partial_z F - \frac{i}{2}\nabla_{\bot}^2 F + \frac{i}{2}\gamma(E_{ext}F - \partial_x\phi F) &= 0 \\
    -\partial_z B - \frac{i}{2}\nabla_{\bot}^2 B + \frac{i}{2}\gamma(E_{ext}B - \partial_x\phi B) &= 0
\end{align}
\end{subequations}
with the nonlinear coupling constant $\gamma = k^2x_0^2n_0^2r_{\text{eff}}$, which contains the refractive index of the unperturbed
crystal $n_0$, the effective element of the electro-optic tensor $r_{\text{eff}}$, and the transverse scaling constant $x_0$.
$\nabla_{\bot}^2$ denotes the transverse Laplacian.
The propagation ($z$-)axis is scaled to the diffraction length $L_D = kx_0^2$. The values used in the simulations
are $x_0 = 10\,\mu\text{m}$ (typical beam widths are about $1\dots
2\,x_0$), $r_{\text{eff}} = 280\,\text{pm/V}$, $n_0 = 2.35$, $E_{ext} = 2.5\,\text{kV/cm}$,
and $k = 2\pi n_0/\lambda$ with $\lambda = 532\,\text{nm}$. The total propagation distance was chosen to be $4L_D$ (11.1~mm).

The results of the simulations are shown in Fig.~\ref{fig:dipole-numerics}.
Image (a) and (b) show the exit face of the dipole
beam at times $t=100~\tau$ and $t=280~\tau$ after the start of the simulation.
Although the
numerical and experimental results differ in detail, the main
features of the interaction are very similar. The splitting of the dipole into
a trapped and non-trapped part can clearly be observed in
Fig.~\ref{fig:dipole-numerics}(a), although it is not as
pronounced as in the experiment. As time elapses the trapped part grows in
intensity drawing energy from the non-trapped part
(Fig.~\ref{fig:dipole-numerics}(b)).
The second main feature, the exceptionally long timescale is reproduced as well,
with the time constant $\tau$ corresponding to approximately $\tau \approx 15~s$.

To summarize, we have shown the existence of a stable counterpropagating
dipole-mode vector soliton in a photorefractive SBN crystal. The
vector soliton differs considerably in many aspects from its counterpart in
copropagating geometry. The timescale of the transient dynamics are
significantly larger. During the formation the beams split up, and a
trapped and non-trapped part of the dipole can be observed.

\end{document}